\documentclass[pra,aps]{revtex4}
\usepackage{epsfig,amsmath,amsfonts}

\begin{document}

\title{Deutsch-Jozsa algorithm for continuous variables} 

\author{Arun K.\ Pati$^{*}$ and Smauel L.\ Braunstein}

\address{School of Informatics, University of Wales, Bangor 
LL57 1UT, United Kingdom}
\address{$^{*}$Institute of Physics, Bhubaneswar-751005, Orissa, INDIA.}


\begin{abstract}
We present an idealized quantum continuous variable analog of the 
Deutsch-Jozsa algorithm which can be implemented on a perfect 
continuous variable quantum computer. Using the Fourier transformation and 
XOR gate appropriate for continuous spectra we show that under ideal
operation to infinite precision that there is an infinite reduction in
number of query calls in this scheme.
\end{abstract}

\maketitle

In principle, quantum computers can have remarkable computational powers
which classical computers cannot \cite{DJCV_sl,sl1}. In the last few years it 
has been shown that it is possible for quantum computers to perform 
certain computational tasks faster than any classical computer 
\cite{pb,DJCV_rf,DJCV_dd,DJCV_dj,bv,drs,div}. 
Quantum computation exploits quantum interference and entanglement
to outperform its classical counterparts. The first algorithm promising
benefits from quantum parallelism was discovered by Deutsch and Jozsa 
\cite{DJCV_dj}. Soon after this Shor discovered \cite{DJCV_ps} his now famous
algorithm for factoring large numbers \cite{ej}. Subsequently, a fast 
quantum search algorithm was discovered by Grover \cite{DJCV_lg,DJCV_lgr}; 
in addition, the time-dependent generalization of Grover's algorithm and 
its stability under unitary perturbation has been studied \cite{DJCV_akp}.

It may be mentioned that all these algorithms are usually designed for 
qubits. However, in nature there are other classes of quantum systems 
whose observables form, for example, continuous spectra. Usually a 
continuous variable can be anything, e.g., position, momentum, 
energy (unbounded), the amplitudes of the electromagnetic field, etc. 
It is important to know how these algorithms can be generalized to
continuous quantum variables. By learning how, we might make progress 
towards discovering new algorithms which are perhaps more naturally
formulated using continuous variables.

Recently continuous quantum information has played an important role in
teleportation \cite{sb} and even error-correction codes \cite{slb,sl2} with
a possible implementation using linear devices \cite{sam}. Moreover, 
quantum computation over continuous variables has also been studied.
It was found that the universal continuous variable quantum computation 
can be effected using simple non-linear operations with coupling provided 
solely by linear operations \cite{DJCV_lb}. 
Just as standard quantum computation 
can be thought of as the coherent manipulation of two-level systems 
qubits, continuous quantum computation can be thought of as the 
manipulation of qunats.

The first algorithm to have been studied for potential implementation
using continuous quantum variables was the Grover virtual database search
\cite{pbl}. Here, we go on to generalize the Deutsch-Jozsa algorithm
to continuous variables. This scheme naively gives {\em an infinite speed-up
over classical function evaluation}.

To start with, let us recall the standard Deutsch-Jozsa algorithm for 
qubits. In this case we are given a number $i\in\{0,\ldots, 2^n-1\} 
\equiv B^n$  
and a ``black box'' or ``oracle query'' that computes a binary function 
$f(i): B^n \rightarrow B$. Further, the function $f(i)$ which only takes 
values $0$ or $1$ is {\it promised \/}
to be either constant or balanced (with an equal number of each type of
outcome over all input strings). The aim is to determine this property for
$f(i)$, i.e., whether it is constant or balanced. On a classical computer 
in the worst case the oracle query requires $O(2^n)$ function evaluations. 
However, if one calculates the function using reversible quantum operations 
then only a single function evaluation is required to achieve the goal 
\cite{DJCV_dj,cemm}.

In the continuous variable setting we pose the problem in the following way.
Suppose there is a particle located somewhere along the $x$-axis. Since
$x$ is a continuous variable it can take value from $-\infty$ to $+ \infty$
(in practice it may be from $-L$ to $L$, where $L$ is some length
scale involved). Suppose there are two persons Alice and Bob playing a
game \cite{mike}.
Alice tells Bob a value of $x$ and Bob calculates some function 
$f(x)$ which takes values $0$ or $1$. Further, Bob has promised Alice
that he will use a function which is either constant or balanced.
A constant function is $0$ or $1$ for all
values of $x \in (-\infty, + \infty)$. 
For a balanced function, $f(x)=0$ or $1$ for exactly half of the cases.
One can define the balanced function more precisely in the following manner. 
Imagine that the interval for the continuous variable $x$ has been 
divided into $n$ sub-intervals. Let $\mu$ be the Lebesgue measure on 
${\mathbb R}$. A function $f(x)$ is balanced provided the Lebesgue measure 
of the support for where the function is zero is identical to the 
Lebesgue measure of the support for where the function is one, i.e., 
$\mu(\{ x \in {\mathbb R}\,| f(x)=0\})
=\mu(\{ x \in {\mathbb R}\, | f(x) =1 \})$.
Now, Alice wants to know whether Bob has chosen a constant or balanced 
function. In the classical scenario since there are an infinite number of 
possiblities for $x$ Alice needs to ask Bob (who has the oracle) an 
infinite number of times!  However, we can show if we use a perfect
continuous variable quantum computer and unitary operators that can be 
implemented on them, then a single function evaluation is required to know 
this global property of the function.

Let us consider a continuous variable system whose Hilbert 
space is infinite dimensional and spanned by a basis state
$\arrowvert x \rangle$ satisfying the orthogonality condition 
$\langle x \arrowvert x' \rangle = \delta(x - x')$. In a continuous variable 
scheme a basic operation is the Fourier transformation between position 
and momentum variables in phase space (analog to the Walsh-Hadamard 
transformation for qubits). We can define the Fourier transformation as 
an active operation on a qunat state $\arrowvert x \rangle$ as
\begin{equation}
{\cal F} \arrowvert x \rangle = \frac{1}{ \sqrt{\pi} } \int dy\; e^{2ixy} 
\arrowvert y \rangle\;,
\label{AKP2_eq}
\end{equation}
where both $x$ and $y$ are in the position basis. This has been used
in developing error correction codes \cite{slb,sam}  
and Grover's algorithm
for continuous variables \cite{pbl}. This Fourier
transformation can be easily applied in physical situations. For example,
when $\arrowvert x \rangle$ represents quadrature eigenstate of a mode of 
the electromagnetic field, ${\cal F} \arrowvert x \rangle$ is simply an
eigenstate of the conjugate quadrature produced by a $\pi/2$ phase delay.

Another useful gate on a continuous variable quantum computer is XOR 
gate (analogous to the controlled NOT gate for qubits but without the cyclic 
condition) defined as \cite{sam} 
\begin{equation}
\arrowvert x \rangle \arrowvert y \rangle \rightarrow
\arrowvert x \rangle \arrowvert x + y \rangle\;.
\end{equation}
Further, we assume that given a classical circuit for computing $f(x)$ there
is a quantum circuit which can compute a unitary transformation $U_f$ on a
continuous variable quantum computer. If a quantum circuit exists that
transforms
\begin{equation}
\arrowvert x \rangle \arrowvert y \rangle \rightarrow
\arrowvert x \rangle \arrowvert y+ f(x) \rangle\;,
\end{equation}
then by linearity it can also act on any superposition of qunat states.
For example, if we evaluate the function on a state (\ref{AKP2_eq}) along 
with another qunat state $\arrowvert z \rangle$, we have
\begin{equation}
{\cal U}_f ( {\cal F} \arrowvert x \rangle \arrowvert z \rangle )
= \frac{1}{\sqrt \pi } \int dy\; e^{2ixy}
\arrowvert y \rangle \arrowvert z+f(y) \rangle\;.
\end{equation}
This shows that using quantum parallelism for idealized qunat computers 
one can evaluate all possible values of a function simultaneously with one
application of ${\cal U}_f$.

Now, we present the Deutsch-Jozsa algorithm for a continuous variable quantum
computer. The set of instructions for deciding the constant or balanced
nature of function $f(x)$ are give below 

(i). Alice stores her query in a qunat register prepared in an ideal position
eigenstate $\arrowvert x_0 \rangle$ and attaches another qunat in a position
eigenstate $\arrowvert \pi/ 2 \rangle$. So the two quants are in the state
$\arrowvert x_0 \rangle \arrowvert \pi/2 \rangle$

(ii). She creates superpositions of qunat states by applying the Fourier 
transformation to the query qunat and the target qunat. The resulting state 
is given by
\begin{equation}
{\cal F} \arrowvert x_0 \rangle {\cal F} \arrowvert \pi/2 \rangle 
= \frac{1}{\pi } \int dx\, dy \,e^{2ix_0 x + i \pi y} 
\arrowvert x \rangle \arrowvert y \rangle\;.
\end{equation}

(iii). Bob evaluates the function using the unitary operator ${\cal U}_f$.
The state transforms as
\begin{equation}
\frac{1}{ \sqrt{\pi} } \int dx \, e^{2ix_0 x + i \pi f(x)} 
\arrowvert x \rangle {\cal F} \arrowvert \pi/2 \rangle \;.
\label{AKP2_eq2}
\end{equation}
Here, the key role is played by the ancilla qunat state 
$\arrowvert \pi/2 \rangle$. To see how the function evaluation takes 
place consider the intermediate steps given by
\begin{eqnarray}
{\cal U}_f (\arrowvert x \rangle {\cal F} \arrowvert {\pi}/{2} \rangle) =
\frac{1}{\sqrt{\pi}} \int dy\, e^{i \pi y} {\cal U}_f 
(\arrowvert x \rangle \arrowvert y \rangle)
= (-1)^{f(x)} \arrowvert x \rangle {\cal F} \arrowvert {\pi}/{2} \rangle\;.
\end{eqnarray}
If the function $f(x)=0$ there is no sign change and if $f(x)=1$ there is
a sign change. After the third step performed by Alice, she has a quant state
in which the result of Bob's function evaluation is encoded in the 
amplitude of the qunat superposition state given in (\ref{AKP2_eq2}). 
To know the nature of the function she now performs an inverse Fourier 
transformation on her qunat state.

(iv). The qunat states after Fourier tranform is given by
\begin{equation}
\arrowvert q \rangle = \frac{1}{\pi } \int dx\, dx' e^{2ix(x_0- x')} 
(-1)^{f(x)} \arrowvert x' \rangle {\cal F} \arrowvert \pi/2 \rangle\;.
\end{equation}

(v). Alice measures her qunat by projecting onto the original position
eigenstate $\arrowvert x_0 \rangle$. In an ideal continuous variable scheme 
the correct projection operator is defined as  \cite{cohen}
\begin{equation}
P_{\Delta x_0} = \int_{x_0- \Delta x_0/2}^{x_0 + \Delta x_0/2}
dy\, \arrowvert y \rangle \langle y \arrowvert\;.
\end{equation}
As has been explained in \cite{pbl,cohen} if the observable has a
continuous spectrum then the measurement cannot be performed precisely
but must involve some spread $\Delta x_0$. Therefore, the action of 
projection onto the qunat state after step (iv) is given by
\begin{equation}
P_{\Delta x_0} \arrowvert q \rangle =
\frac{1}{\pi} \int dx \int_{x_0- \Delta x_0/2}^{x_0 + \Delta x_0/2}
dy\,  e^{2ix(x_0- y)} (-1)^{f(x)} 
\arrowvert y \rangle {\cal F} \arrowvert \pi/2 \rangle\;.
\end{equation}

Now consider two possibilities. If the function is constant then the above
equation reduces to 
$\pm \arrowvert x_0 \rangle {\cal F} \arrowvert \pi/2 \rangle$. 
[In simplifying we need to use the Dirac delta function 
$({1}/{\pi}) \int dx \; e^{2ix(x_0- y)} = \delta(x_0 -y)$.]
This means that if Alice measures $\arrowvert x_0 \rangle$ she is sure
that $f(x)$ is definitely constant. In the other case, i.e., when 
the function is balanced she will not get the measurement outcome to be
$\arrowvert x_0 \rangle$. In fact, in the balanced case the outcome is
orthogonal to the constant case as the result gives zero. 
Therefore, a single function evaluation (follwed by a  measurement onto 
$\arrowvert x_0 \rangle$) in a qunat quantum computer can decide whether the
promised function is constant or balanced. Unlike the qubit case, in the
{\it idealized\/} continuous variable case the {\em reduction in the number 
of query calls is from infinity to one}.

In conclusion, we have generalised the primitive quantum algorithm
(Deutsch-Jozsa algorithm) from the discrete case to the {\it idealized\/}
continuous case. It may be worth mentioning that as in error correction 
codes for continuous-variablees \cite{slb}, if one replaces the 
Hadamard transform and XOR gate by their continuous-variable analogs 
in original Deutsch-Jozsa algorithm for qubit case, then the idealized
algorithm works perfectly.  This theoretically demonstrates the power of 
quantum computers to exploit the superposition principle giving an 
{\em infinite speed up compared to classical scenario}. This idealized analysis
has not considered the affects of finite precision in measurement or
state construction and so whether it may be implemented experimentally 
remains an open question for further study. Part of the difficulty in 
extending this work
in this direction is that defining an oracle for continuous variables
appears to be a difficult task, one that we have carefully avoided here.
An alternate way forward might be to consider some sort of ``hybrid''
approach involving both qunats and qubits. This is precisely what Seth
Lloyd considers in the following chapter.

\vskip 0.2truein

\noindent
AKP thanks P.\ van Loock and R.\ Simon for useful feedback.  AKP also thanks 
G. Giedke for discussions during Benasque Science Center-2000 in Spain
on defining balanced function for continuous variables.

\end{document}